\documentclass[notitlepage,aps,pra,twoside,superscriptaddress,showkeys,nofootinbib]{revtex4-1}
\usepackage{graphicx}
\usepackage{amsmath,amsfonts,amssymb,amsthm,amsbsy,mathtools}
\usepackage{bbold}
\usepackage{epic}
\usepackage{eepic}
\usepackage{mathrsfs}
\usepackage[all]{xy}
\usepackage{lmodern}
\usepackage[T1]{fontenc}
\usepackage[utf8]{inputenc}
\usepackage{color}
\usepackage{xr}
\usepackage{blindtext}
\usepackage[utf8]{inputenc}
\usepackage[T1]{fontenc}
\usepackage{amssymb} \usepackage{color,graphicx} \usepackage{amsmath}
\usepackage{amsbsy} \usepackage{amsthm} 
\usepackage{bm} \usepackage{float} 
\usepackage{placeins,cals}
\usepackage{setspace,ulem}
\usepackage[colorlinks=true,allcolors=blue!80!black]{hyperref}
\usepackage{xfrac}
\usepackage{amsfonts}
\usepackage{latexsym}
\usepackage{amsmath,amssymb}
\usepackage{mathrsfs}
\usepackage{color,xcolor}
\usepackage{slashed}
\usepackage{url}
\usepackage{empheq}
\usepackage{textcomp}
\usepackage{dcolumn}
\usepackage{ulem}
\usepackage{subfigure}
\usepackage{amsfonts}
\usepackage{epsfig}
\usepackage{bbm}
\usepackage{tabularx}
\usepackage{multirow}
\usepackage{mathtools}
\usepackage{xfrac}
\usepackage{epsfig}
\usepackage{amsfonts}
\usepackage{latexsym}
\usepackage{mathrsfs}
\usepackage{hyperref}
\usepackage{setspace}
\usepackage{color}
\usepackage{bm}
\usepackage{orcidlink}
\usepackage{isotope}
\usepackage{braket}
\def \x{\mathbf{ x}}
\def \K{\mathbf{ K}}
\def \k{\mathbf{ k}}
\def \p{\mathbf{ p}}
\def \q{\mathbf{ q}}

\def \Tr{\text{Tr}}

\def\beq{\begin{equation}}
	\def\eeq{\end{equation}}
\def\bea{\begin{eqnarray}}
	\def\eea{\end{eqnarray}}
\usepackage{mhchem}

\def\k{\textbf{k}}
\newcommand{\ketbra}[2]{| #1\rangle\!\langle #2 |}

\usepackage{xcolor} 
\colorlet{BLUE}{blue}

\allowdisplaybreaks
\makeatletter
\usepackage{cancel}
\usepackage{color}

\begin{document}

\title{A Microscopic Approach to Gravitationally Induced Entanglement of Quantum Clocks} 

	\author{M. Zarei\,\orcidlink{0000-0001-7744-2817}}
\email{m.zarei@iut.ac.ir}
\affiliation{Department of Physics, Isfahan University of Technology, 84156-83111 Isfahan, Iran}
\affiliation{Quantum Technology Research Group, Isfahan University of Technology, Isfahan 84156-83111, Iran}

\begin{abstract}

We investigate the dynamics of entanglement generation between two quantum clocks mediated by gravitational interactions within a quantum Boltzmann
equation (QBE) framework. We model each clock as a two level system (characterized by spinor quantum field)
prepared in a coherent superposition of internal energy eigenstates.
We then analyze how gravitational coupling generates quantum correlations between spatially separated clocks. Our
formalism incorporates relativistic time dilation effects together
with graviton mediated interactions. This allows us to study the interplay between internal clock dynamics and weak spacetime curvature at a microscopic level. 
Within this approach we derive the orthogonalization time - the minimal time required for a clock state to evolve into an orthogonal state - and show how gravitational interactions modify this fundamental timescale. 
We find that entanglement arises from energy dependent
phase shifts generated by gravitational potentials and by
graviton mediated forward scattering processes appearing in the QBE.
Our results provide a microscopic description of gravitationally
induced entanglement between quantum clocks and connect quantum field theoretic techniques with recent proposals for probing the quantum nature of gravity using quantum clocks.

\end{abstract}

\maketitle

\section{Introduction}

Quantum mechanics and general relativity constitute the two fundamental
pillars of modern physics. Despite their extraordinary experimental success, their conceptual foundations differ significantly, and a
consistent description of physical systems where both quantum and
gravitational effects cannot be ignored remains an open question. Exploring
the interplay between these theories is expected to provide new
insights into the nature of spacetime and quantum systems
\cite{Bose:2017nin,Marletto:2017kzi,Bose:2023gwh, Marletto:2024ltk, Vicentini:2024noi, Abrahao:2023lle, Arani:2023ljw, Colas:2024xjy, Sugiyama:2024equ, Gallock-Yoshimura:2024wpz, Chevalier:2020uvv, Bassi:2017szd, Kent:2021ljj,bose2017spin,marletto2017gravitationally,Krisnanda:2019glc,Christodoulou2019,danielson2022gravitationally,galley2022no,christodoulou2023locally,Carney:2018ofe, kafri2013noise, Kafri:2014zsa, kafri2015bounds, Bahrami:2015wma, anastopoulos2015probing, bose2017spin, marletto2017gravitationally, haine2021searching, Qvarfort:2018uag, Carlesso:2019cuh, PhysRevA.101.063804, howl2021non, Matsumura:2020law, Pedernales:2021dja, liu2021gravitational, Datta:2021ywm, Trillo:2024kqz}. 
In recent years, quantum clocks - quantum systems possessing internal
degrees of freedom that evolve in time - have emerged as a promising
setup for investigating the interface between quantum mechanics and
gravity \cite{Zych:2011hu,Zych:2012ut,ZychBrukner2015QEEP,Delahamette2023IndefMetric}.
A typical realization involves a massive particle whose internal states
serve as a clock while the particle itself is placed in a coherent
superposition of spatial trajectories. For example, in a Mach-Zehnder
matter wave interferometer located in Earth's gravitational field, the particle that travels along two paths experiences different
gravitational potentials. This causes the internal clock to accumulate
different proper times along each trajectory
\cite{Zych:2011hu,Zych:2012ut}. 
Such systems provide a natural framework
for probing the concept of proper time in quantum mechanics.

According to general relativity, proper time flows at different rates in different regions of spacetime due to gravitational time dilation. This
implies that clocks located at different gravitational potentials tick
at different rates \cite{Hartle2003,Weinberg1972}.
 When a quantum clock
is placed in a spatial superposition across regions with different
gravitational potentials, the accumulated proper time difference can become encoded in its internal state. This feature allows gravitational
time dilation to influence quantum interference phenomena.

The proposed experiments rely on the quantum complementarity principle, which indicates that interference disappears in the conditoin that which path information becomes available. If a clock interferes along two paths with different
proper times, measuring the clock state can in principle reveal the
particle’s path. This information reduces the visibility of the
interference pattern, an effect often referred to as
gravitationally induced decoherence. Several experimental proposals have
been developed to test this effect using interferometric setups
involving massive particles and quantum clocks
\cite{Khosla:2016tss,Roura:2018cfg,HuHuYangLin:2023jgl,Arya:2024hgf}.
 One approach uses optical pulses to generate coherent
superpositions of clock states trapped in spatially separated optical
tweezers. This enables interferometric measurements of gravitational
redshift through oscillations in internal state visibility
\cite{Balatsky:2025GravityDephasing}. Other schemes use optical lattice clocks to investigate manifestations of mass energy equivalence,
including gravitational redshift and also exploring photon mediated interactions that can generate
entanglement and synchronization in arrays of strontium or ytterbium
clocks \cite{Chu:2024dpv}. 
Extensions to distributed clock networks have
also been proposed, where entangled clocks separated by large distances
can perform nonlocal proper time interferometry via photon interference
\cite{Fromonteil:2025ClockNetworks,Covey:2025DistributedClock}. 
These
developments aim to probe regimes in which quantum systems evolve in
classical curved spacetime beyond the Newtonian approximation, while
also enabling tests of foundational aspects of quantum theory in gravitational settings
\cite{Mattei:2025QuantumLimits,Svozil:2025QuantumClocks}.

As both experimental and theoretical investigations increasingly explore
situations where relativistic and quantum effects coexist, it becomes
important to develop frameworks to describe quantum clocks
within a relativistic quantum field theoretic setting. In particular, a
microscopic description of how gravitational interactions influence the
coherence and correlations of quantum clocks is still lacking.

In this work, we study the dynamics of quantum clocks within a density
matrix formalism based on the QBE
\cite{Kosowsky:1994cy,Bavarsad:2009hm,Bartolo:2018igk,Bartolo:2019eac,Hoseinpour:2020hic,Zarei:2021dpb}.
This approach allows us to analyze graviton mediated interactions between
clocks and to determine the conditions under which gravitational
coupling generates entanglement. The framework
combines internal clock dynamics with
relativistic corrections and field theoretic descriptions of gravity,

	in the spirit of the microscopic treatment of spin entanglement
	through effective interactions derived from quantum electrodynamics
	in Ref.~\cite{Zarei:2026rbx},

thereby offering a microscopic perspective on gravitationally induced
correlations between quantum clocks.

The paper is organized as follows. 
In Sec. \ref{clockmodel}, we review the clock model.
In Sec. \ref{clockmodel-QBE}, we use the QBE to rederive the dynamical evolution of quantum clocks and the associated phase shifts arising from gravitational interactions.
In Sec. \ref{clockmodel-GR}, we turn to general relativistic corrections and incorporate time dilation effects into the microscopic framework.
In Sec. \ref{clockmodel-graviton}, we consider the interaction of two spin-$\tfrac{1}{2}$ systems mediated by gravitons and analyze the resulting coupling between the clocks.
In Sec. \ref{clockmodel-entanglement}, we study the entanglement dynamics by computing the negativity of the evolved state.
Finally, in Sec. \ref{conclusion} we conclude.

  \section{The clock model}
  \label{clockmodel}
  Any system in a superposition of energy eigenstates can function as a reference clock to define time evolution. The simplest case considered is a two level system, such as a particle with an internal degree of freedom.
  The clock is assumed to follow a semiclassical and approximately static trajectory. This allows us to neglect the special relativistic effects. However, it can still be located in a gravitational field, like near Earth or near a massive object. Hence, general relativity can still be included as long as spacetime is treated classically and the clock's worldline is well defined.
  The observer need not be colocated with the clock; they may interact with it via a probe system and perform measurements remotely. Only the internal degrees of freedom of the clock are relevant here. The internal Hamiltonian in the clock's rest frame is \cite{CastroRuiz2017,Zych:2011hu,Pikovski:2013qwa,Castro-Ruiz:2019nnl}
  \begin{equation}
  	H_{\text{int}} = E_0 \ketbra{0}{0} + E_1 \ketbra{1}{1}~, \label{Hint0}
  \end{equation}
  with the energy gap defined as $\Delta E = E_1 - E_0 = \hbar\omega > 0$.
  Here $\omega$ is the angular frequency, and $\hbar$ is retained explicitly.
  A convenient operational definition of a unit of time for a quantum clock is given by the orthogonalization time, namely the minimum time required for a state to evolve into another state orthogonal to it. For a two level system, the optimal clock state is the balanced superposition
  \begin{equation}
  	\ket{\psi_{\mathrm{in}}}
  	= \frac{1}{\sqrt{2}} (\ket{0} + \ket{1})~,
  	\label{psi0}
  \end{equation}
  which maximizes sensitivity to phase accumulation. The corresponding optimal measurement basis is
  \(
  \ket{\pm} = (\ket{0} \pm \ket{1})/\sqrt{2}
  \),
  the eigenbasis of the Pauli \(X\) operator.
  To characterize the time evolution of this clock state one can employ the density matrix formalism. The initial state \eqref{psi0} has density matrix
  \begin{equation}
  	\rho(0)
  	= \ket{\psi_{\mathrm{in}}}\bra{\psi_{\mathrm{in}}}
  	= \frac{1}{2}
  	\left(
  	\ket{0}\bra{0}
  	+ \ket{0}\bra{1}
  	+ \ket{1}\bra{0}
  	+ \ket{1}\bra{1}
  	\right)~.
  \end{equation}
  Assuming unitary evolution under the Hamiltonian \eqref{Hint0}, with eigenenergies \(E_0\) and \(E_1\), the state at proper time \(t\) is obtained as
  \begin{equation}
  	\ket{\psi(t)}
  	= \frac{1}{\sqrt{2}}
  	\left( e^{-iE_0 t/\hbar} \ket{0}
  	+ e^{-iE_1 t/\hbar} \ket{1} \right)~,
  \end{equation}
  and the corresponding density matrix becomes
  \begin{equation}
  	\rho(t)
  	= \frac{1}{2}
  	\left(
  	\ket{0}\bra{0}
  	+ e^{i\omega t}\ket{0}\bra{1}
  	+ e^{-i\omega t}\ket{1}\bra{0}
  	+ \ket{1}\bra{1}
  	\right)~.
  \end{equation}
  We quantify the distinguishability between the evolved and initial states by the fidelity as
  \begin{equation}
  	F(t)
  	= \left| \braket{\psi(0)|\psi(t)} \right|^2
  	= \left|
  	\frac{e^{-iE_0 t/\hbar}
  		+ e^{-iE_1 t/\hbar}}{2}
  	\right|^2
  	= \cos^{2}\!\left( \frac{\omega t}{2} \right)~.
  \end{equation}
  Orthogonality is reached when \(F(t)=0\), which occurs whenever
  \begin{equation}
  	\cos \left( \frac{\omega t}{2} \right) = 0~.
  \end{equation}
  The smallest positive solution defines the minimal orthogonalization time
  \begin{equation}
  	t_{\perp}
  	= \frac{\pi\hbar}{\Delta E}~. \label{t1}
  \end{equation}
  The same conclusion follows directly from the density matrix picture. Fidelity also quantifies the closeness between two density matrices. The trace overlap between the initial and time evolved states is
  \begin{equation}
  	\Tr[\rho(0)\rho(t)]
  	= \frac{1}{2}
  	\left( 1 + \cos(\omega t) \right)~.
  \end{equation}
  At \( t = t_{\perp} \), one finds
  \begin{equation}
  	\Tr[\rho(0)\rho(t_{\perp})] = 0~,
  \end{equation}
  confirming that the states are orthogonal.
  More generally, for any pure initial state \( \rho(0) = \ket{\psi(0)}\bra{\psi(0)} \), orthogonality is reached at the smallest \( t>0 \) such that
  \begin{equation}
  	\Tr[\rho(0)\rho(t)] = 0~.
  \end{equation}
  Since for pure states
  \begin{equation}
  	\Tr[\rho(0)\rho(t)]
  	= |\braket{\psi(0)|\psi(t)}|^{2}~,
  \end{equation}
  this condition is precisely equivalent to the orthogonality of the state vectors. For mixed initial states the same trace criterion applies, although exact orthogonality may not be achievable; if it is never reached, there is no finite \(t_{\perp}\).
  
  The orthogonalization time \(t_{\perp}\) is also defined as the minimum time required for a quantum system to evolve from a given state to another state that is orthogonal to it, i.e., a state that is perfectly distinguishable and has zero overlap with the initial one. This notion provides a natural operational time scale for quantum clocks, since the functioning of a clock relies on the ability of a physical system to undergo distinguishable changes during its evolution.
  A fundamental constraint on how rapidly such evolution can occur is given by the Mandelstam-Tamm quantum speed limit \cite{mandelstam1945}. For a quantum system with a time independent Hamiltonian \(H\) and energy uncertainty \(\sigma_E=\sqrt{\langle H^2\rangle-\langle H\rangle^2}\), the time required to reach an orthogonal state satisfies
  \begin{equation}
  	t_{\perp} \geq \frac{\pi \hbar}{2 \sigma_E}~.
  \end{equation}
  This bound shows that the rate of quantum evolution is limited by the spread of energy in the state: larger energy uncertainty allows for faster evolution and therefore shorter orthogonalization times.
  In certain idealized systems this bound can be saturated. For example, in a two level system prepared in an equal superposition of energy eigenstates separated by an energy gap \(\Delta E\), one has \(\sigma_E=\Delta E/2\), and the orthogonalization time becomes equal to \eqref{t1}.
  For more general systems, particularly those involving multiple energy levels or mixed states, the dynamics may not reach perfect orthogonality within this minimal time, and the Mandelstam-Tamm bound provides only a lower limit on the achievable evolution time.
  
  Gravitational effects enter naturally in this context through mass energy equivalence. Since the energy of the clock contributes to the gravitational field it generates, a clock prepared in a superposition of energy eigenstates corresponds to a superposition of gravitational fields when the gravitational interaction is treated quantum mechanically. Consequently, different energy components of the state lead to slightly different gravitational time dilations.
  To illustrate this effect, consider a second clock located at a coordinate distance \(x\) from the first one. Due to gravitational time dilation generated by the energy difference \(\Delta E\) between the clock states, the difference in the proper times accumulated by the second clock in the weak field limit is \cite{CastroRuiz2017,Zych:2011hu,Pikovski:2013qwa}
  \begin{equation}
  		\tau_1-\tau_0 = -\frac{G\,\Delta E}{c^{4} x}\,t
  	~,
  \end{equation}
  where \(\tau_1\) and \(\tau_0\) correspond to the first clock being in the excited state \(\ket{1}\) and ground state \(\ket{0}\), respectively.
  Here \(G\) denotes Newton's gravitational constant, and \(t\) represents the proper time measured by an observer sufficiently far from the clock that its gravitational potential is negligible.

 \begin{figure}[t]
 	\centering
 	\includegraphics[scale=0.3]{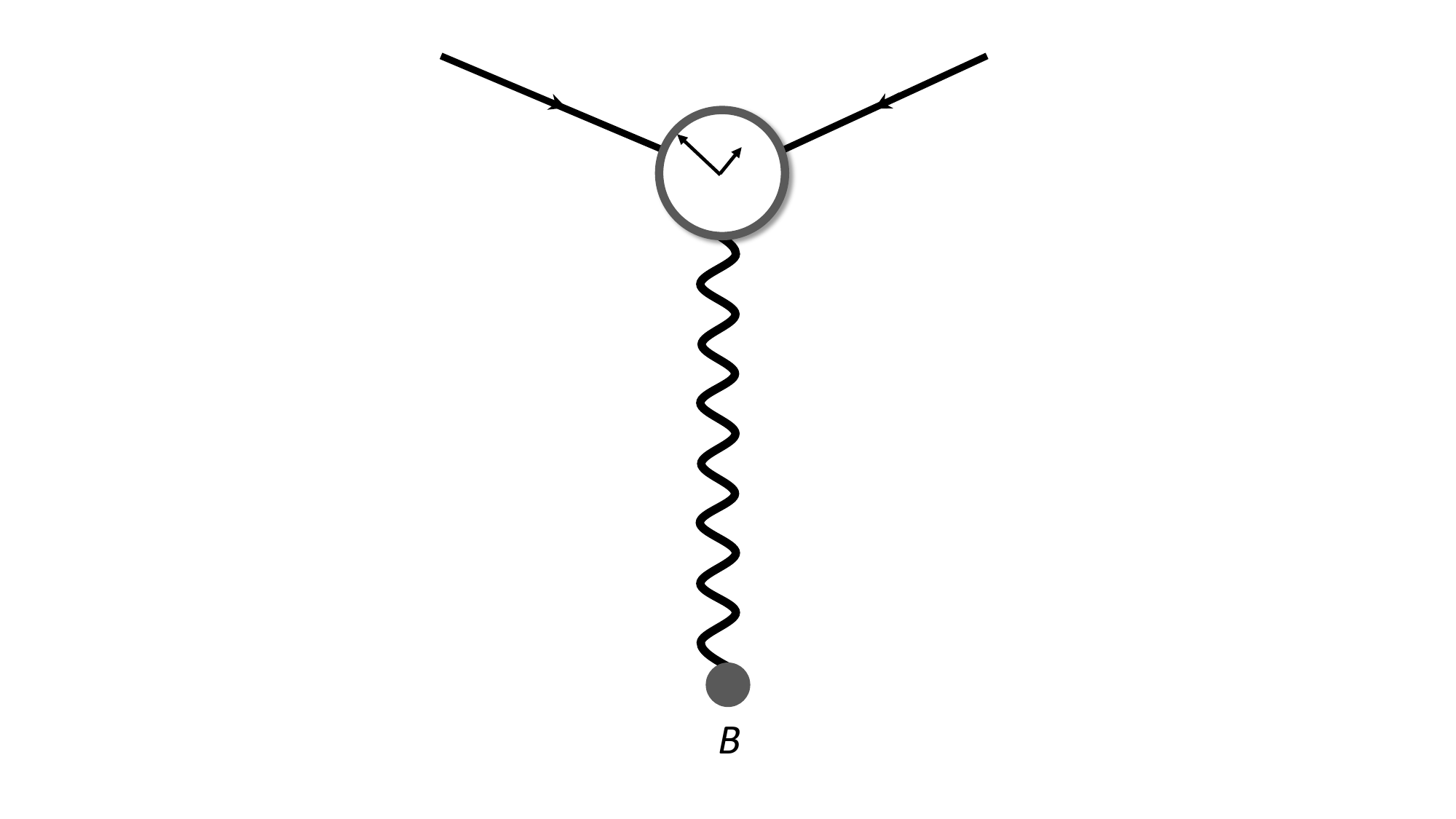}
 	\vspace{-0.5cm}
 	\caption{The quantum clock system of spin $1/2$ particles (solid lines) in the background of magnetic field $B$ (wavy line).}
 	\label{fig1}
 	 	\vspace{-0.4cm}
 \end{figure}

\section{Quantum clocks and the quantum Boltzmann equation}
\label{clockmodel-QBE}

In the following, we investigate the dynamics of quantum clocks from a microscopic perspective using the QBE formalism.
The QBE describes the evolution of a quantum system interacting with its environment~\cite{Kosowsky:1994cy,Bavarsad:2009hm,Bartolo:2018igk,Bartolo:2019eac,Hoseinpour:2020hic,Zarei:2021dpb}
\begin{align}
	\label{QBE0}
	(2\pi)^3\delta^3(0)\dot{\rho}_{ij}(\mathbf{k},t)
	={}&
	\frac{i}{\hbar}
	\left\langle
	\big[
	\hat{H}_{\textrm{int}}(0),
	\hat{\mathcal{D}}_{ij}(\mathbf{k})
	\big]
	\right\rangle
	\nonumber\\
	&-
	\frac{1}{\hbar^2}
	\int_0^t ds\,
	\left\langle
	\Big[
	\hat{H}_{\textrm{int}}(s),
	\big[
	\hat{H}_{\textrm{int}}(0),
	\hat{\mathcal{D}}_{ij}(\mathbf{k})
	\big]
	\Big]
	\right\rangle~,
\end{align}
where $\hat{H}_{\textrm{int}}$ is the interaction Hamiltonian describing the scattering process, $\hat{\mathcal{D}}_{ij}(\mathbf{p})
=a_j^\dagger(\mathbf{p})a_i(\mathbf{p})$ is the number operator associated with the fermion.
In this formalism, the expectation value of an operator $\hat{O}$ is given by
\begin{equation}
	\label{exp1}
	\langle \hat{O}(\mathbf{k}) \rangle
	=
	\mathrm{tr}[\hat{\rho}\hat{O}(\mathbf{k})]
	=
	\int d\mathbf{p}\,
	\left\langle
	\mathbf{p}
	\middle|
	\hat{\rho}\hat{O}(\mathbf{k})
	\middle|
	\mathbf{p}
	\right\rangle,
\end{equation}
with $d\mathbf{p}=d^3p/(2\pi)^3$.
The density operator of the system is defined as
\begin{equation}
	\hat{\rho}=\int d\p\rho_{ij}(\mathbf{p})\hat{\mathcal{D}}_{ij}(\mathbf{p})~, \label{exp1}
\end{equation}
where $\rho_{ij}$ is the density matrix of the clock system.
In Eq.~\eqref{QBE0}, the first term on the right hand side represents forward scattering, where the momenta remain unchanged but phase shifts alter the quantum coherence. The second term describes collision or decay processes, which can induce decoherence.

	In the following current-field coupling, field expansions, and
	Gordon decomposition, we use $\hbar=c=1$, with the electromagnetic
	coupling absorbed into $A_\mu$. We restore $\hbar$ in the Zeeman
	Hamiltonian and subsequent evolution equations.
The interaction between the Dirac field and the background electromagnetic potential is introduced through the current field coupling
\begin{equation}
	\hat{H}_{\mathrm{int}}
	=
	\int d\tau\,d^3x\;
	A_\mu(x)\,
	\bar{\psi}^{-}(x)
	\gamma^\mu
	\psi^{+}(x)
	\,V\,
	\delta^4_{\sigma_0}(x-\bar{x}(\tau))~.
	\label{Hint-0}
\end{equation}
The factor $\delta^4_{\sigma_0}(x-\bar{x}(\tau))$ represents the spatial localization of the matter degrees of freedom around a classical worldline $\bar{x}(\tau)$ with $\tau$ denoting the proper time. This distribution can be written as a Gaussian wavepacket profile multiplied by a temporal delta function
\begin{equation}
	\delta^4_{\sigma_0}(x-\bar{x}(\tau))
	=
	\delta(x^0-\bar{x}^0(\tau))
	\frac{1}{(2\pi\sigma_0^2)^{3/2}}
	\exp\!\left[
	-\frac{|\mathbf{x}-\bar{\mathbf{x}}(\tau)|^2}{2\sigma_0^2}
	\right]~.
\end{equation}
This form models the finite spatial extension of the localized matter system characterized by the width $\sigma_0$.
The positive- and negative-frequency parts of the Dirac field are expanded in momentum space as
\begin{align}
	\psi^{+}(x)
	&=
	\int d\mathbf{k}\,
	\sum_r
	a_r(\mathbf{k})
	u_r(\mathbf{k})
	e^{-ik\cdot x},
	\\
	\bar{\psi}^{-}(x)
	&=
	\int d\mathbf{k}\,
	\sum_r
	a_r^\dagger(\mathbf{k})
	\bar{u}_r(\mathbf{k})
	e^{ik\cdot x}~,
\end{align}
where $d\mathbf{k}\equiv d^3k/(2\pi)^3$, $a_r(\mathbf{k})$ and $a_r^\dagger(\mathbf{k})$ denote the annihilation and creation operators associated with momentum $\mathbf{k}$ and spin index $r$, while $u_r(\mathbf{k})$ are the corresponding Dirac spinors.
To isolate the magnetic interaction responsible for spin dynamics, the Dirac current can be rewritten using the Gordon decomposition
\begin{equation}
	\bar{\psi}\gamma^\mu\psi
	=
	\frac{i}{2m}
	\left(
	\bar{\psi}\partial^\mu\psi
	-
	(\partial^\mu\bar{\psi})\psi
	\right)
	+
	\frac{1}{2m}
	\partial_\nu
	\left(
	\bar{\psi}\Sigma^{\mu\nu}\psi
	\right)~,
\end{equation}
where
$
\Sigma^{\mu\nu}=\frac{i}{2}[\gamma^\mu,\gamma^\nu].
$
The second term gives rise to the spin-magnetic coupling once the derivatives act on the electromagnetic potential.
We focus on the spatial components of the tensor operator, the relation
\begin{equation}
	\Sigma^{ij}
	=
	\epsilon^{ijk}
	\begin{pmatrix}
		\sigma_k & 0\\
		0 & \sigma_k
	\end{pmatrix}~,
\end{equation}
connects the Dirac tensor structure to the Pauli spin matrices.
Using the identity relating derivatives of the vector potential to the magnetic field
$
\partial_i A_j-\partial_j A_i=\epsilon_{ijk}B_k
$
the interaction Hamiltonian reduces to a magnetic dipole coupling between the spin degrees of freedom and the background field.
In the nonrelativistic limit this produces the standard Zeeman interaction
\begin{equation}
	H_{\mathrm{int}}
	=
	-\frac{\gamma\hbar}{2}\,
	\mathbf{B}\cdot\boldsymbol{\sigma}~,
\end{equation}
where $\gamma$ denotes the gyromagnetic ratio. For a magnetic field oriented along the $z$ direction, $\mathbf{B}=B\hat{z}$, the interaction simplifies to
\begin{equation}
	H_{\mathrm{int}}
	=
	-\frac{\hbar\omega_0}{2}\sigma^3~,
\end{equation}
with $\omega_0 = \gamma B$ representing the Larmor frequency.
In the narrow wavepacket limit, the Gaussian factor becomes sharply peaked around 
$\mathbf{q}'=\mathbf{q}$ and can be approximated by a three dimensional Dirac delta 
distribution, enforcing momentum matching in the integral.
Substituting the mode expansion of the fermionic fields into the interaction Hamiltonian \eqref{Hint-0} and performing the integration over $\x$ yields an operator expression of the form
\begin{equation}
	\hat{H}_{\mathrm{int}}
	=
	-\frac{\hbar\omega_0}{2}
	V
	\sum_{rr'}
	\int d\tau\,d\mathbf{q}\,
	\bar{u}_{r'}(\mathbf{q})
	(I\otimes\sigma^3)
	u_r(\mathbf{q})
	\,a_{r'}^\dagger(\mathbf{q})
	a_r(\mathbf{q})\,
	\delta\!\left(x^0-\bar{x}^0(\tau)\right)
	\delta^3_{\sigma_0}\!\left(\bar{\mathbf{x}}(\tau)\right)~,
\end{equation}
where $I$ and $\sigma^3$ denote the identity and the third Pauli matrix respectively and $\delta^3_{\sigma_0}(\bar{\mathbf{x}}(\tau))$ describes the spatially localized wave packet of the clock and accounts for the finite spatial extent of the interaction region.
In the nonrelativistic limit the clock trajectory varies slowly, allowing the approximations
\(
\bar{x}^0(\tau)\simeq\tau
\)
(with $c=1$)
and
\(
\bar{\mathbf{x}}(\tau)\simeq\bar{\mathbf{x}}
\).
After performing the $\tau$ integration and substituting $\hat{H}_{\mathrm{int}}$ into the forward scattering term of the quantum Boltzmann equation \eqref{QBE0}, one obtains
\begin{eqnarray}
	(2\pi)^3\delta^3(0)\,\dot{\rho}_{ij}
	&=&
	-i\frac{\omega_0}{2}
	V\,\delta^3_{\sigma_0}(\bar{\mathbf{x}})
	\int d\mathbf{q}
	\sum_{rr'}
	\bar{u}_{r'}(\mathbf{q})
	(I\!\otimes\!\sigma^3)
	u_r(\mathbf{q})
	\nonumber\\
	&&\times
	\Big\langle
	a^\dagger_{r'}(\mathbf{q})a_r(\mathbf{q})
	a^\dagger_i(\mathbf{k})a_j(\mathbf{k})
	-
	a^\dagger_i(\mathbf{k})a_j(\mathbf{k})
	a^\dagger_{r'}(\mathbf{q})a_r(\mathbf{q})
	\Big\rangle ~.
\end{eqnarray}
Using the canonical anticommutation relations and the definition of the density matrix elements, this expression reduces to
\begin{eqnarray}
	(2\pi)^3\delta^3(0)\,\dot{\rho}_{ij}
	&=&
	-i\frac{\omega_0}{2}
	V\,\delta^3_{\sigma_0}(\bar{\mathbf{x}})
	\int d\mathbf{q}
	\sum_{rr'}
	\bar{u}_{r'}(\mathbf{q})
	(I\!\otimes\!\sigma^3)
	u_r(\mathbf{q})
	\nonumber\\
	&&\times
	(2\pi)^6
	\delta^3(\mathbf{q}-\mathbf{k})
	\delta^3(\mathbf{k}-\mathbf{q})
	\left[
	\delta_{ri}\rho_{r'j}
	-
	\delta_{jr'}\rho_{ir}
	\right]~.
\end{eqnarray}
Performing the momentum integration then gives
\begin{equation}
	(2\pi)^3\delta^3(0)\,\dot{\rho}_{ij}
	=
	-i\frac{\omega_0}{2}
	V\,\delta^3_{\sigma_0}(\bar{\mathbf{x}})
	(2\pi)^3\delta^3(0)
	\sum_{rr'}
	\bar{u}_{r'}(\mathbf{k})
	(I\!\otimes\!\sigma^3)
	u_r(\mathbf{k})
	\left[
	\delta_{ri}\rho_{r'j}
	-
	\delta_{jr'}\rho_{ir}
	\right]~.
\end{equation}
Finally, using the normalization condition
\(
V\,\delta^3_{\sigma_0}(\bar{\mathbf{x}})=1
\),
the evolution equation for the density matrix simplifies to
\begin{equation}
	\dot{\rho}_{ij}
	=
	-i\frac{\omega_0}{2}
	\sum_{rr'}
	\bar{u}_{r'}(\mathbf{k})
	(I\!\otimes\!\sigma^3)
	u_r(\mathbf{k})
	\left[
	\delta_{ri}\rho_{r'j}
	-
	\delta_{jr'}\rho_{ir}
	\right]~. \label{rho_evolution}
\end{equation}
Equation~\eqref{rho_evolution} can be solved using the initial state introduced in Eq.~\eqref{psi0}. The corresponding initial density matrix in the $\{|0\rangle,|1\rangle\}$ basis is
\begin{equation}
	\rho(0)=\frac{1}{2}
	\begin{pmatrix}
		1 & 1\\
		1 & 1
	\end{pmatrix}~,
\end{equation}
which represents an equal superposition of the two energy eigenstates.
Solving Eq.~\eqref{rho_evolution}, one finds that the diagonal elements remain constant while the off diagonal elements acquire a phase determined by the energy splitting. The time evolved density matrix therefore takes the form
\begin{equation}
	\rho(t)=\frac{1}{2}
	\begin{pmatrix}
		1 & e^{i\omega_0t}\\
		e^{-i\omega_0t} & 1
	\end{pmatrix}~.
\end{equation}
To determine the orthogonalization time, we evaluate the trace overlap between the initial and evolved states,
\begin{equation}
	\mathrm{tr}\!\left[\rho(0)\rho(t)\right]
	=
	\frac{1}{2}
	\left(1+\cos\!\left(\omega_0t\right)\right)~.
\end{equation}
Orthogonality is reached when this overlap vanishes, which yields the condition
\begin{equation}
	\cos\!\left(\omega_0t\right)=-1~.
\end{equation}
Consequently, the orthogonalization time is
\begin{equation}
	t_\perp
	=
	\frac{\pi}{\omega_0}
	=
	\frac{\pi\hbar}{\Delta E}~,
\end{equation}
in agreement with the result obtained previously
in Sec.~\ref{clockmodel}.

 \begin{figure}[t]
	\centering
	\includegraphics[scale=0.3]{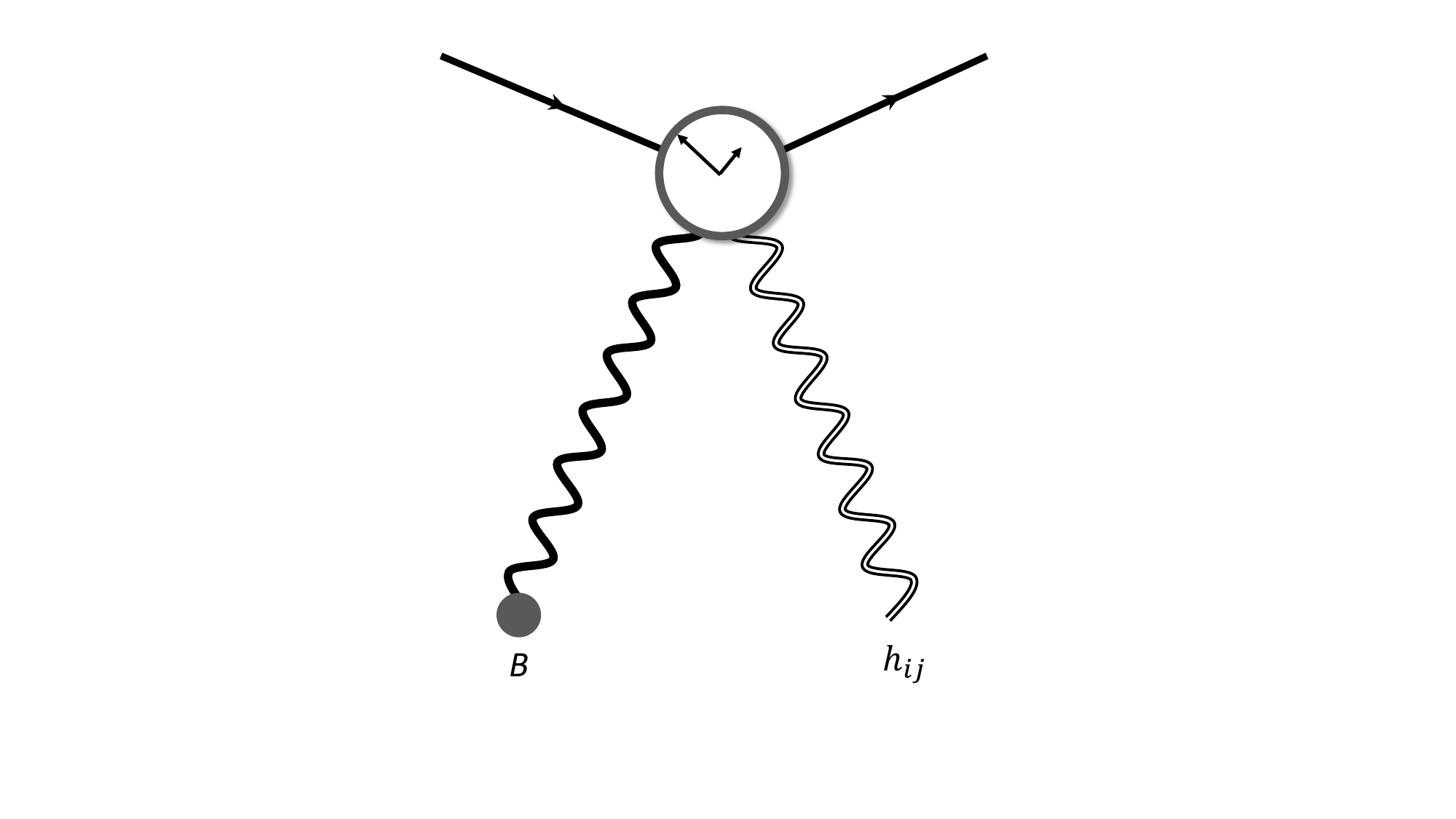}
	\vspace{-0.9cm}
	\caption{The quantum clock system of spin $1/2$ particles (solid lines) in the background of magnetic field $B$ (wavy line) and gravitational field $h_{ij}$ (double wavy line). }
	\label{fig2}
\end{figure}

\section{General relativistic corrections}
\label{clockmodel-GR}

As described earlier, we model the quantum clock as a localized fermionic system evolving in the presence of a weak gravitational perturbation \(h_{\mu\nu}\). We consider a scattering process involving fermionic degrees of freedom interacting with the gravitational field background \(h_{\mu\nu}\) (see Fig.~\ref{fig2}).
	We use $\hbar=c=1$ in the intermediate field theoretic expressions
	and restore $\hbar$ in the effective spin Hamiltonian below.
The corresponding interaction Hamiltonian can be written as
\begin{equation}
	\hat{H}_{\mathrm{int}}
	=
	\frac{\kappa}{2}
	\int d\tau\,d^3x\;
	h^{\mu\nu}(x)\,
	A_\mu(x)\,
	\bar{\psi}^-(x)\gamma_\nu\psi^+(x)\,
	V\,\delta^4_{\sigma_0}\!\left(x-\bar{x}(\tau)\right)~,
	\label{Hmodel1}
\end{equation}
where $\kappa=\sqrt{16\pi G}$.
In the presence of a weak gravitational background, the interaction Hamiltonian
can be made more explicit by isolating the spin contribution. Using the Gordon
decomposition and neglecting spatial gradients of the metric perturbation, the
interaction reduces to
\begin{equation}
	\hat{H}_{\mathrm{int}}
	=
	-\frac{\kappa}{2m}
	\int d\tau\,d^3x\;
	\epsilon^{mal}\,
	\partial_a A^m(x)\,
	h_{ln}(x)\,
	\bar{\psi}^-(x)\,
	(I\!\otimes\!\sigma_n)\,
	\psi^+(x)\;
	V\,\delta^4_{\sigma_0}\!\left(x-\bar{x}(\tau)\right)~.
\end{equation}
We now specialize to a static, weak gravitational field described by a
Newtonian potential. The metric perturbation is taken to be
\begin{equation}
	\kappa h_{ln}(x)
	=
	2\Phi(\mathbf{x})\,\delta_{ln},
	\qquad
	\Phi(\mathbf{x})
	=
	-\frac{GM}{|\mathbf{x}|}~,
\end{equation}
which is valid in the nonrelativistic and weak field limit. Substituting this
form into the interaction Hamiltonian yields
\begin{align}
	\hat{H}_{\mathrm{int}}
	&=
	-\frac{1}{2m}
	\int d\tau\,d^3x\;
	2\Phi(\mathbf{x})\,
	\epsilon^{mal}\,
	\partial_a A^m(x)\,
	\bar{\psi}^-(x)\,
	(I\!\otimes\!\sigma_l)\,
	\psi^+(x)\;
	V\,\delta^4_{\sigma_0}\!\left(x-\bar{x}(\tau)\right)
	\nonumber\\[4pt]
	&=
	-\frac{1}{2m}
	\int d\tau\,d^3x\;
	2\Phi(\mathbf{x})\,
	B^l(x)\,
	\bar{\psi}^-(x)\,
	(I\!\otimes\!\sigma_l)\,
	\psi^+(x)\;
	V\,\delta\!\left(x^0-\bar{x}^0(\tau)\right)
	\delta^3_{\sigma_0}\!\left(\mathbf{x}-\bar{\mathbf{x}}(\tau)\right)~,
\end{align}
where we have identified
$B^l(x)=\epsilon^{mal}\partial_a A^m(x)$ as the magnetic field.

For a magnetic field aligned along the $z$ direction, the interaction further
simplifies according to $B^l\sigma_l=B\sigma^3$. Substituting the mode
expansion of the fermionic fields then leads to the
gravitational correction to the
Zeeman interaction Hamiltonian
\begin{align}
	\hat{H}_{\mathrm{int}}
	=
	-\frac{VB}{2m}
	\sum_{r,r'}
	\int d\tau\,d^3x\,d\mathbf{q}\,d\mathbf{q}'
	\;2\Phi(\mathbf{x})\,
	\delta\!\left(x^0-\bar{x}^0(\tau)\right)\,
	\delta^3_{\sigma_0}\!\left(\mathbf{x}-\bar{\mathbf{x}}(\tau)\right)
	\nonumber\\
	\times\,
	\bar{u}_{r'}(\mathbf{q}')
	(I\!\otimes\!\sigma^3)
	u_r(\mathbf{q})\,
	e^{\,i(q'-q)\cdot x}\,
	a^\dagger_{r'}(\mathbf{q}')
	a_r(\mathbf{q})~.
	\label{Hint-expanded}
\end{align}
Using the delta function $\delta\!\left(x^{0}-\bar{x}^{0}(\tau)\right)$, the
integration over the proper time $\tau$ can be performed explicitly. In the
low energy, no recoil regime considered here, only forward scattering
processes contribute, corresponding to $r'=r$ and
$\mathbf{q}'=\mathbf{q}$. As a result, Eq.~\eqref{Hint-expanded} reduces to
\begin{equation}
	\hat{H}_{\mathrm{int}}
	=
	-\frac{VB}{2m}
	\sum_r
	\int d^3\mathbf{x}\,d\mathbf{q}\;
	2\Phi(\mathbf{x})\,
	\delta^3_{\sigma_0}\!\left(\mathbf{x}-\bar{\mathbf{x}}\right)\,
	\bar{u}_r(\mathbf{q})
	(I\!\otimes\!\sigma^3)
	u_r(\mathbf{q})\;
	a^\dagger_r(\mathbf{q})a_r(\mathbf{q})~,
	\label{Hint-norecoil}
\end{equation}
where we have made explicit that
the retained contribution
does not change either the spin or the momentum of the fermionic modes.
The remaining spatial integral corresponds to the convolution of the
Newtonian potential with the Gaussian profile describing the spatial
localization of the clock. This convolution can be evaluated analytically and
yields
\begin{equation}
	\int d^3\mathbf{x}\;
	\frac{1}{|\mathbf{x}|}\,
	\delta^3_{\sigma_0}\!\left(\mathbf{x}-\bar{\mathbf{x}}\right)
	=
	\frac{1}{|\bar{\mathbf{x}}|}\,
	\operatorname{erf}\!\left(
	\frac{|\bar{\mathbf{x}}|}{\sqrt{2}\sigma_0}
	\right)~,
\end{equation}
where $\operatorname{erf}(x)$ denotes the error function.
We parameterize the effective interaction Hamiltonian, restoring
	$\hbar$, as
\begin{equation}
	\hat{H}_{\mathrm{int}}
	=
	-\frac{\hbar\omega_G}{2}
	\sum_r
	\int d\mathbf{q}\;
	\bar{u}_r(\mathbf{q})
	(I\!\otimes\!\sigma^3)
	u_r(\mathbf{q})\,
	a^\dagger_r(\mathbf{q})a_r(\mathbf{q})~,
	\label{Hint-final}
\end{equation}
which corresponds to a gravitational correction to the
Zeeman splitting.
The proposed expression for the effective frequency,
	in natural units, is
\begin{equation}
	\omega_G
	=
	\frac{B}{m}\,
	\frac{GM}{|\bar{\mathbf{x}}|}
	\operatorname{erf}\!\left(
	\frac{|\bar{\mathbf{x}}|}{\sqrt{2}\sigma_0}
	\right)~.
	\label{omega-def}
\end{equation}

The forward scattering contribution to the QBE \eqref{QBE0}
generated by the effective Hamiltonian~\eqref{Hint-final} governs the
evolution of the reduced spin density matrix $\rho_{ij}$.
Keeping only the forward scattering processes, the QBE takes the form
\begin{equation}
	\dot{\rho}_{ij}
	=
	-i\frac{\omega_G}{2}
	\sum_{r r'}\int d^3\mathbf{q}\;
	\bar{u}_{r'}(\mathbf{q})
	(I\!\otimes\!\sigma^{3})
	u_r(\mathbf{q})
	\left(
	\delta_{ri}\rho_{r'j}
	-
	\delta_{jr'}\rho_{ir}
	\right)~,
	\label{QBE3}
\end{equation}
where $i,j,r,r'$ denote spin indices.
Using the nonrelativistic relation for normalized spinors
\begin{equation}
	\bar{u}_{r'}(\mathbf{k})
	(I\!\otimes\!\sigma^3)
	u_r(\mathbf{k})
	\simeq
	(\sigma^3)_{r'r}~,
\end{equation}
the spinor structure reduces to the matrix elements of the Pauli operator
$\sigma^3$. The QBE then simplifies to the compact operator form
\begin{equation}
	\dot{\rho}
	=
	\frac{i\omega_G}{2}\,
	[\sigma^3,\rho]~,
	\label{unitary-precession}
\end{equation}
which describes unitary spin precession generated by the effective Hamiltonian
$H_{\mathrm{eff}}=-\hbar\omega_G\sigma^3/2$.
To illustrate the resulting dynamics, consider the initial state
\begin{equation}
	\rho(0)
	=
	\frac{1}{2}
	\begin{pmatrix}
		1 & 1\\[2mm]
		1 & 1
	\end{pmatrix}~,
\end{equation}
corresponding to a coherent superposition of the two spin eigenstates.
Solving Eq.~\eqref{unitary-precession} yields
\begin{equation}
	\rho(t)
	=
	\frac{1}{2}
	\begin{pmatrix}
		1 & e^{i\omega_G t}\\[2mm]
		e^{-i\omega_G t} & 1
	\end{pmatrix}~.
	\label{rho-of-t}
\end{equation}

The overlap between the initial and time evolved states is
\begin{equation}
	\mathrm{tr}\!\left[\rho(0)\rho(t)\right]
	=
	\frac{1}{2}\left(1+\cos\omega_G t\right).
\end{equation}
This quantity vanishes when the states become orthogonal, which occurs for
$\omega_Gt_\perp=\pi$.
The orthogonalization time is therefore
\begin{equation}
	t_\perp
	=
	\frac{\pi}{\omega_G}
	=
	\frac{\pi\hbar}{\Delta E_G}~,
	\label{orthotime}
\end{equation}
where $\Delta E_G=\hbar\omega_G$ denotes the gravitationally corrected
energy splitting of the clock.

Equation~\eqref{orthotime} coincides with the result obtained previously \cite{Zych:2011hu,Zych:2012ut}
from the macroscopic description of the clock, demonstrating consistency
between the microscopic QBE dynamics and the effective description in terms
of a gravitationally shifted level splitting.
In the weak field limit, the gravitational potential
$\Phi=-GM/x$ induces a small correction to the rate at which the clock
accumulates phase. To first order,
the elapsed proper time of a stationary clock is
\begin{equation}
		\tau=t+\Delta t
		=
		t\left(1+\frac{\Phi}{c^2}\right)
		=
		t\left(1-\frac{GM}{c^2x}\right)
	~,
\end{equation}
which relates the microscopic energy splitting of the clock to the
gravitational redshift of its ticking rate.

  \begin{figure}[t]
 	\centering
 	\includegraphics[scale=0.3]{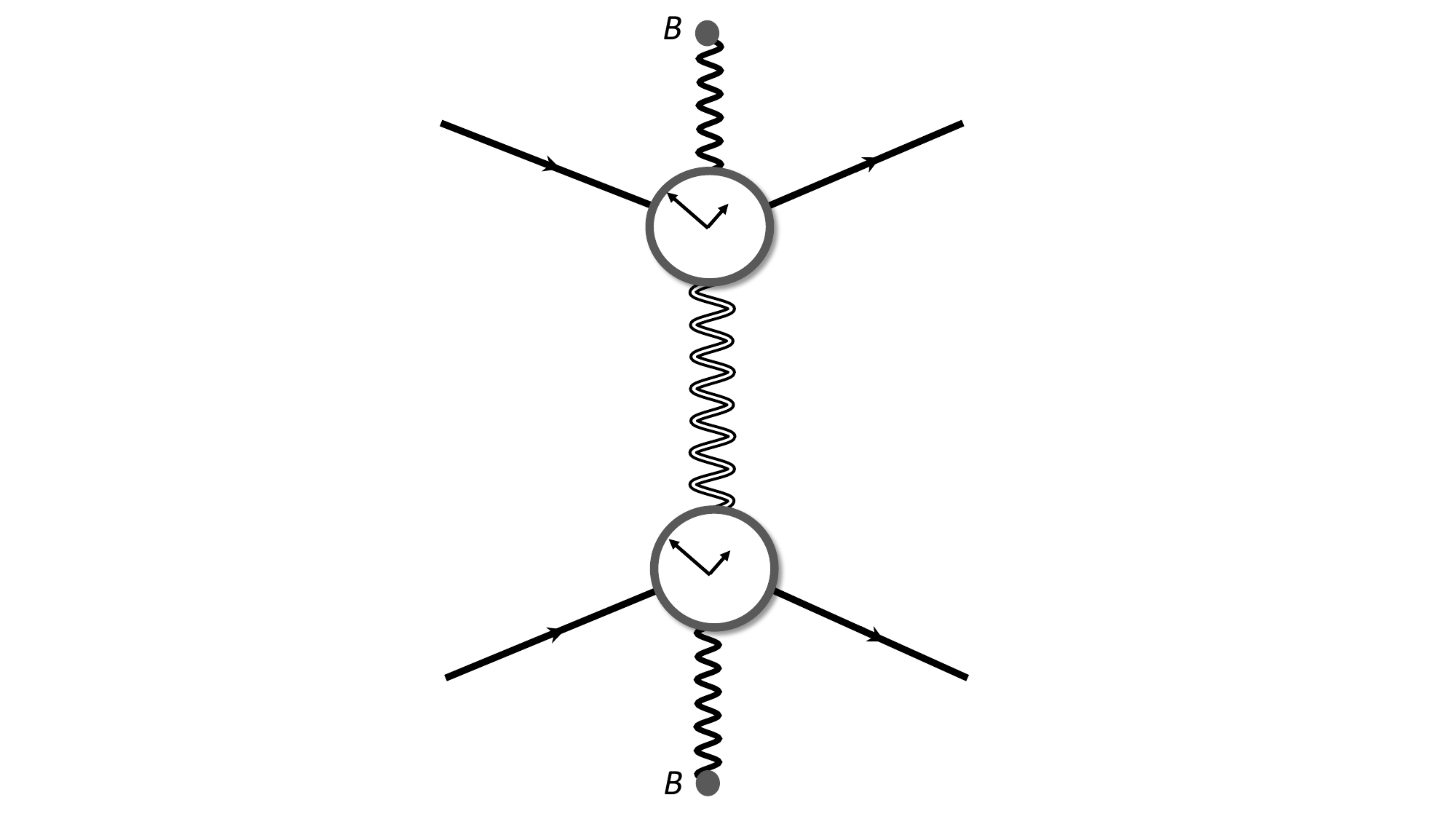}
 	\vspace{0.4cm}
 	\caption{ The interaction of two quantum clock systems (solid lines) trhough graviton (double wavy line). Clocks are in the background of magnetic field $B$ (wavy line). }
 	\label{fig3}
 \end{figure}

 \section{Interaction of two spin-1/2 systems via graviton}
 \label{clockmodel-graviton}
 
 Here, we use an extension of the QBE formalism developed to investigate
 entanglement dynamics between two quantum clock systems, $A$ and $B$,
 providing a generalized framework for their correlated evolution.
 To track the evolution of entanglement, we apply a modified QBE to the
 $4\times4$ density matrix $\rho_{IJ}$ \cite{zarei}
 \begin{align}
 	\label{QBE1}
 	\left[(2\pi)^3\delta^3(0)\right]^2
 	\dot{\rho}_{IJ}(\mathbf{k},t)
 	=
 	\frac{i}{\hbar}
 	\left\langle
 	\big[
 	\hat{H}_{\textrm{int}}(0),
 	\hat{\mathcal{D}}_{IJ}(\mathbf{k})
 	\big]
 	\right\rangle~,
 \end{align}
 where the capital indices are $I,J=1,\dots,4$ and
 $\hat{\mathcal{D}}_{IJ}$ is the number operator.
 Exploiting the independence of the spin Hilbert spaces of the two
 localized systems, the number operator can be written as
  \begin{equation}
\hat{\mathcal D}_{IJ}(\mathbf k)
\simeq
a_i^{A\dagger}(\mathbf k)a_j^A(\mathbf k)
\otimes
a_k^{B\dagger}(\mathbf k)a_l^B(\mathbf k)~,
 \end{equation}
 where the density matrix indices have been folded such that
 \(I\equiv (i,k)\), \(J\equiv (j,l)\) with \(i,j,k,l=1,2\).
 The density operator is defined as
 \begin{equation}
 	\hat{\rho}=\int d\p\rho_{IJ}(\mathbf{p})\hat{\mathcal{D}}_{IJ}(\mathbf{p})~.
 \end{equation}
 Again, in Eq.~\eqref{QBE1}, the term on the right hand side represents forward scattering where the momenta remain unchanged but phase shifts alter the quantum coherence. Here, we put the focus on the forward scattering contribution to the entanglement dynamics.
 	We use $\hbar=c=1$ in the intermediate field-theoretic expressions
 	below and restore physical units in the effective two clock dynamics.

 We consider the process in which two fermions interact via graviton
 exchange, whose dynamics is described by the following effective Hamiltonian
 \begin{eqnarray}
 	\hat{H}_{\textrm{int}}
 	&=&
 	\frac{V^2\kappa^2}{4m^2}
 	\int d\tau\,d\tau'\,d^3x\,d^4x'\;
 	B_i
 	\bar{\psi}^{-}(x')
 	(I\otimes\sigma_j)
 	\psi^{+}(x')
 	\delta_{\sigma'_0}^4(x'-\bar{x}'(\tau'))
 	D^{iji'j'}(x-x')
 	\nonumber\\
 	&&\times
 	B_{i'}
 	\bar{\psi}^{-}(x)
 	(I\otimes\sigma_{j'})
 	\psi^{+}(x)
 	\delta_{\sigma_0}^4(x-\bar{x}(\tau))~,
 	\label{Hmodel-graviton}
 \end{eqnarray}
 where the graviton propagator in the harmonic (de Donder) gauge
 in momentum space is given by~\cite{Bartolo:2018igk}
 \begin{align}
 	D^{\mu\nu\mu'\nu'}
 	=
 	\frac{-1}{K^2}
 	\left(
 	\eta^{\mu\mu'}\eta^{\nu\nu'}
 	+
 	\eta^{\mu\nu'}\eta^{\nu\mu'}
 	-
 	\eta^{\mu\nu}\eta^{\mu'\nu'}
 	\right)~,
 \end{align}
 where $\eta^{\mu\nu}$ is the Minkowski metric tensor and
 $K$ is the exchanged graviton momentum.
We express the effective interaction Hamiltonian in Fourier space as follows
 \begin{align}
 	\hat{H}_{\mathrm{int}}
 	={}&
 	\frac{V^2B^2\kappa^2}{4m^2}
 	\sum_{rr',ss'}
 	\int d^3x\,d^3x'\,
 	d\mathbf{p}\,d\mathbf{p}'\,d\mathbf{q}\,d\mathbf{q}'\,
 	d\mathbf{K}\;
 	\frac{1}{|\mathbf{K}|^2}
 	\delta^3_{\sigma_0}(\mathbf{x}-\bar{\mathbf{x}})
 		\delta^3_{\sigma'_0}(\mathbf{x}'-\bar{\mathbf{x}}') 	
 	e^{-i(\mathbf{K}+\mathbf{p}-\mathbf{p}')\cdot\mathbf{x}}
 	e^{-i(\mathbf{q}-\mathbf{q}'-\mathbf{K})\cdot\mathbf{x}'}
 	\nonumber\\
 	&\times
 	\bar{u}_{r'}(\mathbf{p}')
 	(I\otimes\sigma^3)u_r(\mathbf{p})\,
 	\bar{u}_{s'}(\mathbf{q}')
 	(I\otimes\sigma^3)u_s(\mathbf{q})\,
 		a_{r'}^{A\dagger}(\mathbf{p}')a_r^A(\mathbf{p})
 		a_{s'}^{B\dagger}(\mathbf{q}')a_s^B(\mathbf{q})
 	~.
 	\label{effH-Fourier}
 \end{align}
 	Here $d\mathbf{K}=d^3K/(2\pi)^3$.
  We now integrate over $\mathbf{x}$ and $\mathbf{x}'$
 \begin{align}
 	\hat{H}_{\mathrm{int}}
 	={}&
 	\frac{V^2B^2\kappa^2}{4m^2}
 	\sum_{rr',ss'}
 	\int
 	d\mathbf{p}\,d\mathbf{p}'\,d\mathbf{q}\,d\mathbf{q}'\,
 	d\mathbf{K}\;
 	\frac{1}{|\mathbf{K}|^2}
 	e^{-i(\mathbf{K}+\mathbf{p}-\mathbf{p}')\cdot\bar{\mathbf{x}}}
 	e^{-\sigma_0^2|\mathbf{K}+\mathbf{p}-\mathbf{p}'|^2/2}
 	e^{-i(\mathbf{q}-\mathbf{q}'-\mathbf{K})\cdot\bar{\mathbf{x}}'}
 		e^{-\sigma_0'^2|\mathbf{q}-\mathbf{q}'-\mathbf{K}|^2/2}
 	\nonumber\\
 	&\times
 	\bar{u}_{r'}(\mathbf{p}')
 	(I\otimes\sigma^3)u_r(\mathbf{p})\,
 	\bar{u}_{s'}(\mathbf{q}')
 	(I\otimes\sigma^3)u_s(\mathbf{q})\,
 	a_{r'}^{A\dagger}(\mathbf{p}')a_r^A(\mathbf{p})
 	a_{s'}^{B\dagger}(\mathbf{q}')a_s^B(\mathbf{q})~.
 	\label{effH5}
 \end{align}
 
 To investigate entanglement generation,
 we now proceed by substituting the interaction Hamiltonian
 \eqref{effH5} into the forward scattering term of the QBE
 \eqref{QBE1}. This substitution enables us to determine the evolution
 of $\rho_{IJ}$ up to the second order in $\kappa$.
 	We use $\hbar=c=1$ in the following intermediate expressions,
 	with $\rho_{ik,jl}=\langle i,k|\rho|j,l\rangle$.
 
 By using the integrations over $\mathbf{x}$ and
 $\mathbf{x}'$, we arrive at
 \begin{align}
 	[(2\pi)^3\delta^3(0)]^2\dot{\rho}_{IJ}
 	&=
 -	i\frac{V^2B^2\kappa^2}{4m^2}
 	\sum_{rr'}\sum_{ss'}
 	\int d\p\,d\p'\,d\q\,d\q'\,d\K\,
 	\frac{1}{\K^2}
 	e^{-i(\K+\p-\p')\cdot\bar{\x}'}
 		e^{-\sigma_0'^2|\p'-\p-\K|^2/2}
 	e^{i(\q'-\q+\K)\cdot\bar{\x}}
 	e^{-\sigma_0^2|\q'-\q+\K|^2/2}
 	\notag\\
 	&\quad\times
 	\bar{u}_{s'}(\p')(I\otimes\sigma^3)u_s(\p)\,
 	\bar{u}_{r'}(\q')(I\otimes\sigma^3)u_r(\q)
 	\notag\\
 	&\quad\times
 	\Big[
 	\langle
 	a_{r'}^{A\dag}(\q')a^A_r(\q)
 	b_{s'}^{B\dag}(\p')b^B_s(\p)
 	\hat{\mathcal{D}}_{IJ}(\k)
 	\rangle
 	-
 	\langle
 	\hat{\mathcal{D}}_{IJ}(\k)
 	a_{r'}^{A\dag}(\q')a^A_r(\q)
 	b_{s'}^{B\dag}(\p')b^B_s(\p)
 	\rangle
 	\Big]~.
 	\label{QBE3-1}
 \end{align}
 The wave packet form of particles with finite widths $\sigma_0$ and
 $\sigma'_0$ introduces an effective ultraviolet cutoff for
 $|\mathbf{K}|$
 in the momentum diagonal contribution.
 Substituting $\hat{\mathcal{D}}_{IJ}(\k)$ into Eq.~\eqref{QBE3-1},
 we arrive at
 \begin{align}
 	[(2\pi)^3\delta^3(0)]^2
 	\dot{\rho}_{IJ}
 	&=
 	-\frac{iV^2B^2\kappa^2}{4m^2}
 	\sum_{rr'ss'}
 	\int d\p\,d\p'\,d\q\,d\q'\,d\K\,
 	\frac{1}{\K^2}
 	e^{-i(\K+\p-\p')\cdot\bar{\x}}
 	e^{-i(\q-\q'-\K)\cdot\bar{\x}'}
 	\nonumber\\
 	&\qquad\times
 	e^{-\sigma_0^2|\K+\p-\p'|^2/2}
 	e^{-\sigma_0'^2|\q-\q'-\K|^2/2}
 	\bar{u}_{s'}(\q')(I\otimes\sigma^3)u_s(\q)\;
 	\bar{u}_{r'}(\p')(I\otimes\sigma^3)u_r(\p)
 	\nonumber\\
 	&\qquad\times
 		\Big[
 		\left\langle
 		a_{r'}^\dag(\p')a_r(\p)
 		a_j^\dag(\k)a_i(\k)\right\rangle^A	\left\langle
 		b_{s'}^\dag(\q')b_s(\q)
 		b_l^\dag(\k)b_k(\k)
 		\right\rangle^B
 	\nonumber\\
 	&\qquad\quad
 		-
 		\left\langle
 		a_j^\dag(\k)a_i(\k)
 		a_{r'}^\dag(\p')a_r(\p) 	\right\rangle^A	\left\langle
 		b_l^\dag(\k)b_k(\k)
 		b_{s'}^\dag(\q')b_s(\q)
 		\right\rangle^B
 		\Big]
 	~.
 \end{align}
The expectation values in the above equation are evaluated using the general approximation
 \begin{eqnarray}
 	\langle
 	c_{r'}^\dag(\p')c_r(\p)
 	c_{s'}^\dag(\q')c_s(\q)
 	\rangle
 	&\simeq&
 	(2\pi)^6
 	\delta^3_{\sigma_0}(\p-\q')
 	\delta^3_{\sigma_0}(\q-\p')
 	\delta_{rs'}
 	\rho_{sr'}(\q)~,
 \end{eqnarray}
 following
 \cite{Kosowsky:1994cy,Bavarsad:2009hm,Bartolo:2018igk,
 	Bartolo:2019eac,Hoseinpour:2020hic,Zarei:2021dpb}.
  Integrating over the momenta $\p$, $\p'$, $\q$ and $\q'$
then yields
 \begin{eqnarray}
 	\dot{\rho}_{IJ}
 	&=&
 	-\frac{iV^2B^2\kappa^2}{4m^2}
 	\sum_{rr'ss'}
 	\int d\K\,
 	\frac{1}{|\K|^2}
 	e^{i\K\cdot(\bar{\x}-\bar{\x}')}
 	e^{-\sigma_0^2|\K|^2}
 	\,
 	\bar{u}_{r'}(\k)(I\otimes\sigma^3)u_r(\k)
 	\,
 	\bar{u}_{s'}(\k)(I\otimes\sigma^3)u_s(\k)
 	\notag\\
 	&&\times
 		\Big[
 		\delta_{ir'}\delta_{ks'}\rho_{rs,jl}(\k)
 		-
 		\delta_{rj}\delta_{sl}\rho_{ik,r's'}(\k)
 		\Big]
 	\nonumber\\
 	&=&
 	-\frac{iV^2B^2\kappa^2}{4m^2}
 	\int d\K\,
 	\frac{1}{|\K|^2}
 	e^{i\K\cdot(\bar{\x}-\bar{\x}')}
 	e^{-\sigma_0^2|\K|^2}
 	\Big[
 	\sum_{r's'}
 		\bar{u}_{i}(\k)(I\otimes\sigma^3)u_{r'}(\k)
 		\bar{u}_{k}(\k)(I\otimes\sigma^3)u_{s'}(\k)
 		\rho_{r's',jl}(\k)
 	\notag\\
 	&&\qquad\qquad
 	-
 	\sum_{rs}
 		\bar{u}_{r}(\k)(I\otimes\sigma^3)u_j(\k)
 		\bar{u}_{s}(\k)(I\otimes\sigma^3)u_l(\k)
 		\rho_{ik,rs}(\k) 	
 	\Big]~.
 \end{eqnarray}
 After performing the integration over $\K$
 we obtain
 \begin{eqnarray}
 	\dot{\rho}_{IJ}
 	&=&
 	-\frac{iV^2B^2\kappa^2}{16\pi m^2}
 	\frac{
 		\mathrm{erf}\!\left(|\bar{\x}-\bar{\x}'|/2\sigma_0\right)}
 	{|\bar{\x}-\bar{\x}'|}
 	\Big[
 	\sum_{r's'}
 		\bar{u}_{i}(\k)(I\otimes\sigma^3)u_{r'}(\k)
 		\bar{u}_{k}(\k)(I\otimes\sigma^3)u_{s'}(\k)
 		\rho_{r's',jl}(\k)
 	\notag\\
 	&&\qquad
 	-
 	\sum_{rs}
 		\bar{u}_{r}(\k)(I\otimes\sigma^3)u_j(\k)
 		\bar{u}_{s}(\k)(I\otimes\sigma^3)u_l(\k)
 		\rho_{ik,rs}(\k)
 	\Big]~.
 	\label{rhoIJ}
 \end{eqnarray}
Here, the interaction induces mixing between the components of $\rho^A$ and $\rho^B$,
which leads to the generation of quantum entanglement between the two clocks.
Equation~\eqref{rhoIJ} can be rewritten in the compact form
 \begin{equation}
 	\dot{\rho}_{IJ}
 	=
 	-i\,\frac{V^2B^2\kappa^2}{16\pi m^2}\,
 	\frac{
 		\mathrm{erf}\!\left(|\bar{\x}-\bar{\x}'|/2\sigma_0\right)}
 	{|\bar{\x}-\bar{\x}'|}
 	\,F(i,j,k,l)~.
 \end{equation}
 The function $F$ collects the spinor contractions with folded indices.
 The composite indices follow the tensor product mapping
 $I=2(i-1)+k$ and $J=2(j-1)+l$.
 The function $F$ is defined as
 \begin{eqnarray}
 	F(i,j,k,l)
 	&=&
 	\sum_{r's'}
 		\bar{u}_{i}(\k)(I\otimes\sigma^3)u_{r'}(\k)
 		\bar{u}_{k}(\k)(I\otimes\sigma^3)u_{s'}(\k)
 		\rho_{r's',jl}(\k)
 	\nonumber\\
 	&&\qquad
 	-
 	\sum_{rs}
 		\bar{u}_{r}(\k)(I\otimes\sigma^3)u_j(\k)
 		\bar{u}_{s}(\k)(I\otimes\sigma^3)u_l(\k)
 		\rho_{ik,rs}(\k)
 	~.
 	\label{ttdynamics}
 \end{eqnarray}
 The indices appearing on the right hand side are related to the matrix
 indices $I$ and $J$ on the left hand side through the tensor product
 rule defined above.
 
 This behavior is the signature of an effective energy energy coupling
 between the clocks. Indeed, in the weak field limit gravitational time
 dilation leads to an interaction Hamiltonian of the form
 \cite{CastroRuiz2017}
 \begin{equation}
 	H_{\mathrm{int}}
 	=
 	-\frac{G}{rc^4}H_{A_1}H_{A_2}~,
 	\label{HBA}
 \end{equation}
 where $H_{A_1}$ and $H_{A_2}$ denote the internal Hamiltonians of the
 two clocks separated by a distance $r$. Such a coupling was derived
 and analyzed in Ref.~\cite{CastroRuiz2017}, where it was shown that
 the gravitational interaction generates state dependent phases
 that entangle the clocks.

 	To obtain the joint evolution, we use Eq.~\eqref{HBA}, with
 	$H_{A_1}=H_{A_2}=-\hbar\omega_0\sigma^3/2$ for identical clocks.
 	For the Gaussian profiles, define
 \begin{equation}
 		\epsilon(r)
 		=
 		\frac{G\hbar}{rc^4}
 		\operatorname{erf}\!\left(\frac{r}{2\sigma_0}\right),
 		\qquad
 		g(r)=\frac{\epsilon(r)\omega_0^2}{4}
 	~.
 \end{equation}
 	The parameter $\epsilon$ has units of time, while $g$ has units
 	of angular frequency. 
 	The parameter $\epsilon$ characterizes the gravitational coupling
 	between the clocks separated by the distance
 $r=|\bar{\mathbf{x}}-\bar{\mathbf{x}}'|$.
 		For $r\gg\sigma_0$,
 	$\epsilon=G\hbar/(rc^4)$.
 	The total Hamiltonian is therefore
 \begin{equation}
 		H_{\mathrm{tot}}
 		=
 		-\frac{\hbar\omega_0}{2}
 		\left(\sigma^3\otimes I+I\otimes\sigma^3\right)
 		-\hbar g\,\sigma^3\otimes\sigma^3
 	~.
 \end{equation}
  	This effective model corresponds to $J=-g$; its coefficient
 	must be independently matched to the microscopic calculation.

 The $\rho_{IJ}$ matrix elements are then obtained by solving the following
 joint differential equations:
 \begin{align}
 		\dot{\rho}_{11}
 		=
 		\dot{\rho}_{22}
 		=
 		\dot{\rho}_{33}
 		=
 		\dot{\rho}_{44}
 		=
 		\dot{\rho}_{23}
 		=
 		\dot{\rho}_{32}
 	&=0,
 	\\
 	\dot{\rho}_{12}
 	&=i(\omega_0+2g)\rho_{12},
 	&
 	\dot{\rho}_{13}
 	&=i(\omega_0+2g)\rho_{13},
 	\\
 	\dot{\rho}_{24}
 	&=i(\omega_0-2g)\rho_{24},
 	&
 	\dot{\rho}_{34}
 	&=i(\omega_0-2g)\rho_{34},
 	\\
 	\dot{\rho}_{14}
 	&=2i\omega_0\rho_{14},
 	&
 	\dot{\rho}_{JI}
 	&=(\dot{\rho}_{IJ})^*.
 \end{align}
 	For the initial product state
 	$|+\rangle_A|+\rangle_B$, where
 	$|+\rangle=(|0\rangle+|1\rangle)/\sqrt{2}$, 
 the full solution is given by the following matrix
 \begin{equation}
 		\rho_{[IJ]}(t)
 		=
 		\frac{1}{4}
 		\begin{pmatrix}
 			1
 			& e^{i(\omega_0+2g)t}
 			& e^{i(\omega_0+2g)t}
 			& e^{2i\omega_0t}
 			\\
 			e^{-i(\omega_0+2g)t}
 			& 1
 			& 1
 			& e^{i(\omega_0-2g)t}
 			\\
 			e^{-i(\omega_0+2g)t}
 			& 1
 			& 1
 			& e^{i(\omega_0-2g)t}
 			\\
 			e^{-2i\omega_0t}
 			& e^{-i(\omega_0-2g)t}
 			& e^{-i(\omega_0-2g)t}
 			& 1
 		\end{pmatrix}
 	~.
 	\label{rhoIJ4}
 \end{equation}
 The structure of the off diagonal elements reveals that the interaction
 modifies the phase accumulation rates of the clock coherences.
 In particular, the coherences acquire frequency shifts proportional
 to $\epsilon\omega_0^2$, indicating that the internal
 energy of one clock affects the evolution of the other through the
 gravitational field. As a result, the phase evolution depends on the
 joint energy configuration of the two clock system.
 
 The phase factors appearing in Eq.~\eqref{rhoIJ4} exhibit precisely
 this feature: the dynamical phases depend on the energies of both
 subsystems simultaneously. Consequently, even when the clocks start
 in a separable superposition state, the gravitationally induced phase
 evolution can generate entanglement between the two
 internal degrees of freedom.
 
 	This effective evolution illustrates the mechanism proposed in
 	Ref.~\cite{CastroRuiz2017}. Establishing its microscopic coefficient
 	requires completing the normalization and matching of the
 	graviton exchange calculation.

 \section{Negativity of the evolved state}
 \label{clockmodel-entanglement}
 
 We now quantify the entanglement generated between the two clocks by
 computing the negativity of the evolved density matrix.
 This provides a direct measure of the
 entanglement induced by the gravitational interaction.
 In the framework of gravitationally coupled quantum clocks discussed
 in Ref.~\cite{CastroRuiz2017}, such correlations originate from the
 energy-energy coupling produced by gravitational time dilation.
 
 The dynamical density matrix obtained in the previous section is
 \begin{equation}
 		\rho(t)=\frac{1}{4}
 		\begin{pmatrix}
 			1 & e^{i\alpha^+t} & e^{i\alpha^+t} & e^{2i\omega_0t} \\
 			e^{-i\alpha^+t} & 1 & 1 & e^{i\alpha^-t} \\
 			e^{-i\alpha^+t} & 1 & 1 & e^{i\alpha^-t} \\
 			e^{-2i\omega_0t} & e^{-i\alpha^-t} & e^{-i\alpha^-t} & 1
 		\end{pmatrix}
 	~,
 	\label{eq:rho-full}
 \end{equation}
 where the gravitational interaction modifies the phase accumulation
 rates of the clock coherences through the shifted frequencies
 \begin{equation}
 		\alpha^\pm=\omega_0\pm2g
 	~.
 	\label{eq:alpha-beta-def}
 \end{equation}
 The appearance of the shifted frequencies $\alpha^\pm$ reflects the
 fact that the internal energy of one clock influences the rate of
 evolution of the other. This structure is consistent with the effective
 interaction Hamiltonian \eqref{HBA} derived in
 Ref.~\cite{CastroRuiz2017}, which describes the coupling of the internal
 Hamiltonians of two clocks through gravitational time dilation.
 
 To quantify this entanglement we compute the negativity
 \cite{Horodecki:2009zz}
 \begin{equation}
 	\mathcal{N}(\rho)=\frac{\|\rho^{T_B}\|_1-1}{2}~,
 \end{equation}
 where $\rho^{T_B}$ denotes the partial transpose with respect to the
 second clock. Performing the partial transpose of
 Eq.~\eqref{eq:rho-full} yields
 \begin{equation}
 		\rho^{T_B}(t)=\frac{1}{4}
 		\begin{pmatrix}
 			1 & e^{-i\alpha^+t} & e^{i\alpha^+t} & 1 \\
 			e^{i\alpha^+t} & 1 & e^{2i\omega_0t} & e^{i\alpha^-t} \\
 			e^{-i\alpha^+t} & e^{-2i\omega_0t} & 1 & e^{-i\alpha^-t} \\
 			1 & e^{-i\alpha^-t} & e^{i\alpha^-t} & 1
 		\end{pmatrix}
 	~.
 \end{equation}
 	The eigenvalues of $\rho^{T_B}$ are therefore
 \begin{equation}
 		\lambda_{1,2}
 		=
 		\frac{1\pm\cos(2gt)}{2},
 		\qquad
 		\lambda_\pm
 		=
 		\pm\frac{|\sin(2gt)|}{2}
 	~.
 	\label{eq:lambda-pm}
 \end{equation}
 Only $\lambda_-$ can become negative; therefore the negativity is
 \begin{equation}
 		\mathcal{N}(t)
 		=
 		\frac{1}{2}|\sin(2gt)|
 		=
 		\frac{1}{2}
 		\left|
 		\sin\!\left(\frac{\epsilon(r)\omega_0^2t}{2}\right)
 		\right|
 	~.
 	\label{eq:N-final}
 \end{equation}
 
 	In particular, $\mathcal{N}(0)=0$, as required for the initial
 	product state, and $\mathcal{N}(t)=0$ when the interaction vanishes.
 	Since $|\sin(2gt)|\leq1$, the negativity satisfies
 \begin{equation}
 		0\leq\mathcal{N}(t)\leq\frac{1}{2}
 	~.
 \end{equation}
 	For nonzero $g$, the upper bound is reached when
 \begin{equation}
 		2|g|t=\frac{\pi}{2}+n\pi,
 		\qquad n=0,1,2,\ldots
 	~.
 \end{equation}
 At these instants the negativity reaches
\begin{equation}
		\mathcal{N}_{\max}=\frac{1}{2}
	~.
	\label{eq:N-max}
\end{equation}
Thus the two clock system can reach the maximal entanglement
compatible with a two qubit pure state.

 		In this ideal unitary model, the gravitational coupling determines
 		the entanglement timescale, with the first maximum at
 		$t=\pi/(4|g|)$.
 		For any nonzero coupling the maximal negativity is $1/2$,
 		provided coherence is maintained over this time.
 		This effective model result illustrates the mechanism described
 		in Ref.~\cite{CastroRuiz2017}, where gravity induces entanglement
 		between quantum clocks through energy-dependent phase evolution.

 \section{Conclusion}
  \label{conclusion}
 In this work we investigated the dynamics of quantum clocks interacting
 through gravity within the framework of the QBE. We Model each clock as a two level system prepared in a coherent
 superposition of internal energy eigenstates. We then analyzed how
 gravitational interactions modify the evolution of quantum coherence
 and generate entanglement between spatially separated clocks.
 
 We first examined the microscopic evolution of a single clock in a weak
 gravitational background. Using a Gaussian smeared interaction
 Hamiltonian and performing the nonrelativistic reduction of the
 qubit-gravity coupling, we obtained an effective spin precession
 equation describing the gravitationally induced phase evolution of the
 clock state. Within the QBE formalism, this dynamics naturally leads to a
 quantum speed limit governed by the gravitationally shifted frequency,
 showing explicitly how gravitational time dilation modifies the
 orthogonalisation time of a quantum clock.
 We then extended the framework to a pair of clocks interacting through
 graviton exchange. In this microscopic description the dominant
 contribution arises from forward scattering processes mediated by the
 gravitational field. The resulting effective dynamics generates
 energy dependent phase shifts in the joint density matrix of the two
 clocks. This structure reproduces, at the dynamical level, the
 energy-energy coupling
 \begin{equation}
 	H_{\mathrm{int}} = -\frac{G}{r c^{4}} H_A H_B ,
 \end{equation}
 which has been identified as the fundamental mechanism behind
 gravitationally induced clock entanglement
 \cite{CastroRuiz2017}. In our formulation the same mechanism emerges
 from the QBE description of graviton mediated interactions, thereby
 providing a microscopic field theoretic realization of the gravitational
 entanglement mechanism.
 
 The explicit solution of the two clock density matrix allowed us to
analyze the entanglement dynamics in detail. We computed the negativity
of the evolved state and showed that
	in the ideal unitary effective model, a nonzero coupling allows the
	initial product superposition to evolve periodically into a maximally
	entangled two qubit pure state, with negativity $\mathcal{N}=1/2$.
	The coupling determines the entanglement timescale rather than the
	maximal negativity attainable in this model; without the interaction,
	the initially separable state remains separable.
This result illustrates how gravitational time dilation
can affect the dynamical buildup of quantum
correlations.

Several extensions of this framework are possible. One natural
direction is the study of multi clock networks, where collective
gravitational interactions may lead to complex entanglement structures
relevant for distributed quantum sensing and relativistic quantum
metrology. Another important avenue is the inclusion of curved
background geometries beyond the weak field approximation, allowing the
investigation of quantum clock correlations in strong gravitational
fields. Finally, a systematic treatment of gravitational decoherence
within the QBE framework could provide quantitative predictions for
future experiments involving quantum clocks, interferometers, and
long distance quantum communication in gravitational environments.

\section{Acknowledgements}
MZ would like to thank department of Physics and Astronomy “G. Galilei” at University of Padova for warm hospitality while this work was done. He would also like to thank M. Abdi for very useful discussions and comments.

\bibliography{refs.bib}

\appendix

\end{document}